\documentclass[%
 amsmath,amssymb,superscriptaddress,reprint,
prb
]{revtex4-1}

\usepackage{graphicx,subfigure}
\usepackage{dcolumn}
\usepackage{bm}
\usepackage{amsthm}

\usepackage{nicefrac}
\usepackage{etex}                  
\usepackage{pgfplots}
\usepackage{tikz}
\usepackage{tikz-3dplot}
\usepgfplotslibrary{fillbetween}      
\pgfplotsset{compat=newest}
\pgfplotsset{plot coordinates/math parser=false}
\usetikzlibrary{calc,fadings,shapes.misc,3d,arrows,
 decorations,decorations.pathmorphing,
 decorations.text,shapes.geometric,intersections,
 decorations.markings,patterns,spy,mindmap}

\begin{document}

\title{Ultra thin anti-reflective coatings designed using Differential Evolution}

\author{Emmanuel Centeno}
\email{emmanuel.centeno@uca.fr}
\affiliation{Universit\'e Clermont Auvergne, CNRS, SIGMA Clermont, Institut Pascal, BP 10448, F-63000 CLERMONT-FERRAND, FRANCE}

\author{Amira Farahoui}
\affiliation{Universit\'e Clermont Auvergne, CNRS, SIGMA Clermont, Institut Pascal, BP 10448, F-63000 CLERMONT-FERRAND, FRANCE}
\affiliation{Universit\'e Clermont Auvergne, CNRS, Institut de Chimie de Clermont-Ferrand, 63000 Clermont-Ferrand, France}

\author{Rafik Smaali}
\affiliation{Universit\'e Clermont Auvergne, CNRS, SIGMA Clermont, Institut Pascal, BP 10448, F-63000 CLERMONT-FERRAND, FRANCE}

\author{Ang\'elique Bousquet}
\affiliation{Universit\'e Clermont Auvergne, CNRS, Institut de Chimie de Clermont-Ferrand, 63000 Clermont-Ferrand, France}

\author{Fran\c{c}ois R\'everet}
\affiliation{Universit\'e Clermont Auvergne, CNRS, SIGMA Clermont, Institut Pascal, BP 10448, F-63000 CLERMONT-FERRAND, FRANCE}

\author{Olivier Teytaud}
\affiliation{Facebook AI Research, Paris, France}

\author{Antoine Moreau}%
\affiliation{Universit\'e Clermont Auvergne, CNRS, SIGMA Clermont, Institut Pascal, BP 10448, F-63000 CLERMONT-FERRAND, FRANCE}

\begin{abstract}
We use a state-of-the-art optimization algorithm combined with a careful methodology to find optimal anti-reflective coatings. Our results show that ultra thin structures (less than $300 \,nm$ thick) outperform much thicker gradual patterns as well as traditional interferential anti-reflective coatings. These optimal designs actually combine a gradual increase of the refractive index with patterns meant to leverage interferential effects. Contrarily to gradual patterns, they do not require extremely low refractive index materials, so that they can actually be fabricated. Using a cheap and easily deployable vapor deposition technique, we fabricated a 4-layer anti-reflective coating, which proved very efficient over the whole visible spectrum despite a total thickness of only 200 nm.
\end{abstract}

\maketitle
The idea of using a gradual medium for suppressing wave reflections between two distinct materials has been proposed as early as 1879 by Lord Rayleigh.\cite{Rayleigh} Properly designed gradual refractive index materials lower unwanted light reflection for a  broad range of frequency and present a large angular acceptance. Since collecting a maximal amount of light is a crucial issue for many devices, like solar cells or detectors for instance, a lot of efforts have been devoted to the realization of such anti-reflective coatings based on the aforementioned graded refractive index (RI) approach. \cite{Raut:2011dz} Original concepts and formalisms have been specifically developed to identify  continuous gradual RI profiles   that would be able to minimize light reflection over a broad range of wavelength and for any incident angle. \cite{Dobrowolski:2004wd, Chen:2007p4391, Kim:2013in, Loncar:2013, Good:2016iy, southwell1983} All these RI gradient profiles, described by Gaussian, Quintic or Exponential functions, require two main conditions: (i) the graded layer must start with the refractive indices of the initial medium (often air) and terminate with the RI of the substrate ($n_s$), (ii) an adiabatic transition is necessary, which means that the overall thickness of the AR coating ends up being larger than one wavelength. Both requirements make it however extremely challenging to realize such a thick multilayer whose RI should typically vary from 1 to $n_s$. Two main strategies relying either on the step stair RI approximation or on patterned surfaces are employed to satisfy as far as possible such conditions\cite{Raut:2011dz}. 

The former approach allowed to realize a very low refractive index of $1.05$ using an oblique-angle deposition called GLAncing Angle Deposition method (GLAD) on $SiO_2$ and about $500 \,nm$ thick graded AR coating involving 5-7 layers have been realized\cite{Xi,Kuo:2008p4393,Schubert:2008uj,Cai:2014cu, Chhajed:2011bf}. However technological limitations hinder the realization of thicker graded profiles that integrate hundreds of layers, so that the overall thickness of the already fabricated coatings are smaller than the wavelength -- and their performances are far from the performances of theoretical gradual structures. 

It has been recently underlined that the strategy based on leveraging destructive interference to lower the reflectance may present better performances since the refractive index matching and the adiabatic conditions are not satisfied for realistic devices coating\cite{Schubert:2010dl}. The most simple structure for this interferential approach is the well-known quarter-wave layer which cancels out the reflection for a targeted wavelength and for a given incident angle. Multilayers combining a few quarter-wave and half-wave layers allow to extend these properties over a wider range of frequencies and of incidence angles. This approach leads to very thick structures (hundreds of microns) which are, due to manufacturing constraints, difficult to fabricate\cite{Tikhonravov:1996uj, Boudet} and costly. From the mathematical point of view it is difficult to find the best choice of layers since no analytic solutions can be found for such complex systems. Optimization methods such as Needle \cite{Tikhonravov} or simulated annealing \cite{Boudet} and others\cite{verly2018design} are thus employed to numerically tackle this problem.  The major issue here concerns the convergence of the optimization scheme toward  the best solution. In practice, it amounts  to finding a global minimum for a  cost function in a parameter landscape which presents numerous local minima and except in a few cases\cite{barry} it is not possible to certify that the obtained solution is optimal. We underline that parallel search techniques like genetic algorithms and evolution strategies have some built-in safeguards to forestall misconvergence. \cite{Schubert:2008uj, Poxson:2009tj}. Finally, interferential multilayered coatings often involve a limited number of films whose thicknesses are typically of a few hundreds of nanometers \cite{Schubert:2010dl}. Up to date, the conception of efficient and thin AR coating wobbles between continuously graded RI and interferential strategies.
  
In this work, we use an evolutionary algorithm called Differential Evolution (DE) that is particularly suited for optimizing photonic structures\cite{barry} to find optimal AR coatings whose overall thickness only in the $100-200 \, nm$ range, which roughly corresponds to the thickness of a quarter-wave layer. By realizing a systematic study of the optimal design as a function of the thickness, we show that our optimal AR coatings combining gradual and interferential patterns outperform structures that are more than a micron thick. In addition, our designs satisfy the fabrication constraints since they do not require to synthesize a front layer with a very low refractive index close to $1$. We indeed found feasible and robust structures whose the refractive index ranges from $1.2$ to $n_s=3.5$. As a proof of concept, we have fabricated a $200 \, nm$ thick nano-patterned AR coating with a vapor deposition technique -- a low cost and easily scalable industrial process.

To identify the optimal RI profile for a given thickness $d$, the structure is divided in $N$ sub-layers, named pixels, of equal thickness $d_{pix}=d/N$. An incoming plane wave propagating in air of optical index $n_a=1$ illuminates the AR coating backed by a substrate presenting a high refractive index $n_s=3.5$.  We define the cost function as the  reflection coefficient $<R(\lambda,\theta)>$ averaged for both polarized TE and TM waves  in the wavelength range  $\Delta \lambda=[400, 900] \, nm$  and for an incident angle range $\Delta\theta=[0^{\circ}$, 80$^{\circ}]$: 
\begin{equation}
<R(\lambda, \theta)>=\frac{1}{\Delta \lambda \Delta \theta}   \iint  R(\lambda,\theta)d\lambda d\theta
\end{equation}

We assume that the materials are not dispersive. The numerical calculations are made with an open source code based on the S-matrix method for stability reasons.\cite{Defrance:2016co}. The AR coating  RI profile is encoded in a vector $\textbf x$ whose components are the refractive indices $n_i$ of each pixel $i$ in the range of $[1,N]$. The first and $N^{th}$ layers are respectively adjacent to the air medium and the substrate. We then use Differential Evolution in its current-to-best DE2 version to minimize the cost function.\cite{Storn:1997ul, Price} In this framework, each  generation is constituted of $N_p$ vectors $\textbf x$. The cost function of each vector $\textbf x_n$ $(n=1,2,...,N_p)$  is compared to an offspring $\textbf v$ that combines the difference between two  randomly chosen individuals ($\textbf x_{r1}$ and $\textbf x_{r2}$) and the difference between the best individual $\textbf x_b$ for this generation and the vector $\textbf x_n$:
\begin{equation}
\textbf v= \textbf x_n + \alpha(\textbf x_b-\textbf x_n )+ F(\textbf x_{r1}-\textbf x_{r2}).
\end{equation}
The vector corresponding to the lowest cost function is kept, while the other one is discarded. To increase the diversity of the population, a mutation process with variable-wise mutation rate $CR$ mixes the genes of the parent and of the offspring: the parent  $\textbf x_n$ is replaced by the mutant offspring if the latter lowers the cost function.  Our optimizations are performed with  a population $N_p=30$, a mutation probability $CR=0.8$, $\alpha=0.6$ and $F=0.9$. 
 
 To check that such an approach allows to converge when the number of pixels is increased, the average reflectivity of an AR coating is computed for an increasing number of layers while keeping the overall thickness to $200 \, nm$, Fig. 1. We underline that light propagates from left to right i.e. propagating from air towards the silicon substrate. Without any AR coating (zero pixel) the average reflectivity of the barre silicon is around $41\%$. The reflectivity decreases to $3.3\%$ when the number of pixels is increased. This optimal reflectivity is essentially reached when the number of pixels is larger than 3 -- which corresponds to a nano-structuration at the scale of few tens of nanometer, far below the conventional quarter-wave thickness. Even if the complexity of the optimization problem increases with the number of pixels, the fact that the reflectivity converges towards a minimum value makes us confident that the profiles are close enough to the optimal solution. Furthermore, we have applied this methodology to several thicknesses to make sure that the process had actually converged. This point is extremely important because there is, in general, no way of determining whether a solution produced by a numerical optimization process is optimal. 
 
  The evolution of the RI profile brings an interesting physical insight about what is the best strategy to adopt for a given number of layers. In the case of a small number of sub-domains, 2 or 3 pixels, the optimal AR coating resembles a gradual RI profile (inset of Fig. 1). However, when the number of pixels increases this gradual profile becomes chopped. These RI  discontinuities are a clear signature of an interferential structure that is combined with the graded profile. We assist to the emergence of a stable RI profile as the number of pixels increases. For example, the first  discontinuity arising at a distance of about $70 \, nm$ from the air interface persists when the RI profile is refined. These results also show that any optimal structure with a relatively low number of layers starting with only 4 layers includes a nanostructured pattern for interferential reasons.

\begin{figure}
\centering
\includegraphics[width=\linewidth]{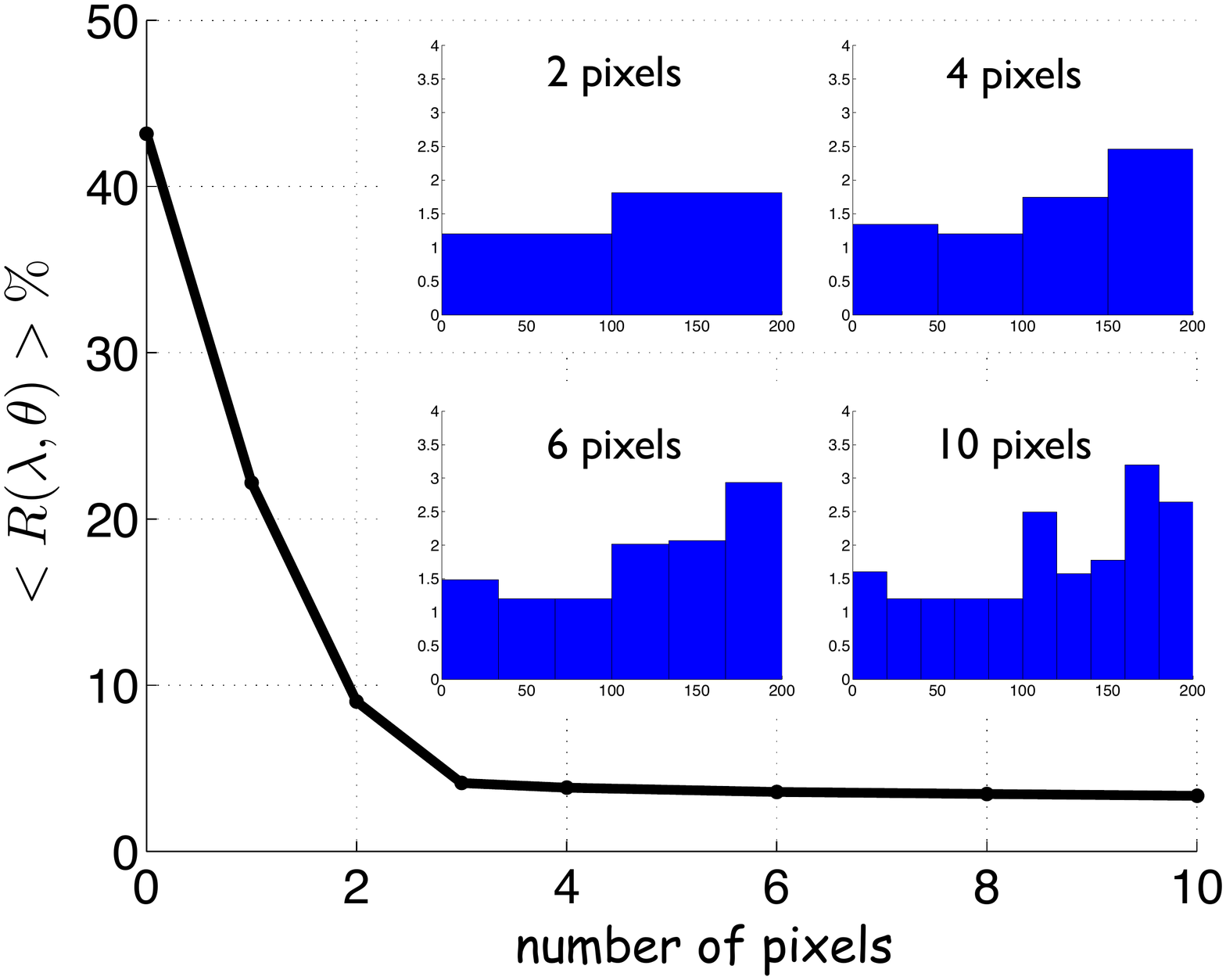}
\caption{Average reflectivity as a function of the number of pixels for a $200 \, nm$ thick AR coating. The insets represent the RI of the sub-layer for an increasing number of pixels. The incoming wave propagates from air to the substrate which are respectively placed  at the left and right sides of the AR coating.}
\end{figure}

We now monitor the AR coating performance as a function of its thickness by calculating the average reflectivity as a function of $d$, Fig. 2. Simulations have been obtained for pixels presenting a thickness $d_{pix}$ settled between $15$ to $20 \, nm$.  Whatever the thickness of the AR coating, the structures produced by the DE2 algorithm outperforms the Gaussian, exponential and quintic graded profiles. Even thick gradual profiles of $1 \, \mu m$ present worth performances  since the refractive index matching condition is not satisfied when a 1.4 minimal threshold is imposed for the optical index. We also observe that thin AR coatings of only $200-400 \, nm $ thick are very efficient since the reflectivity lowers to about $-30 \, dB$, respectively $3.4\%$ and $2.4 \%$. In comparison, the average reflectivity attains $-33 \, dB$ ($2.3 \%$) for a thickness of $1 \, \mu m$.    

\begin{figure}
\centering
\includegraphics[width=9cm]{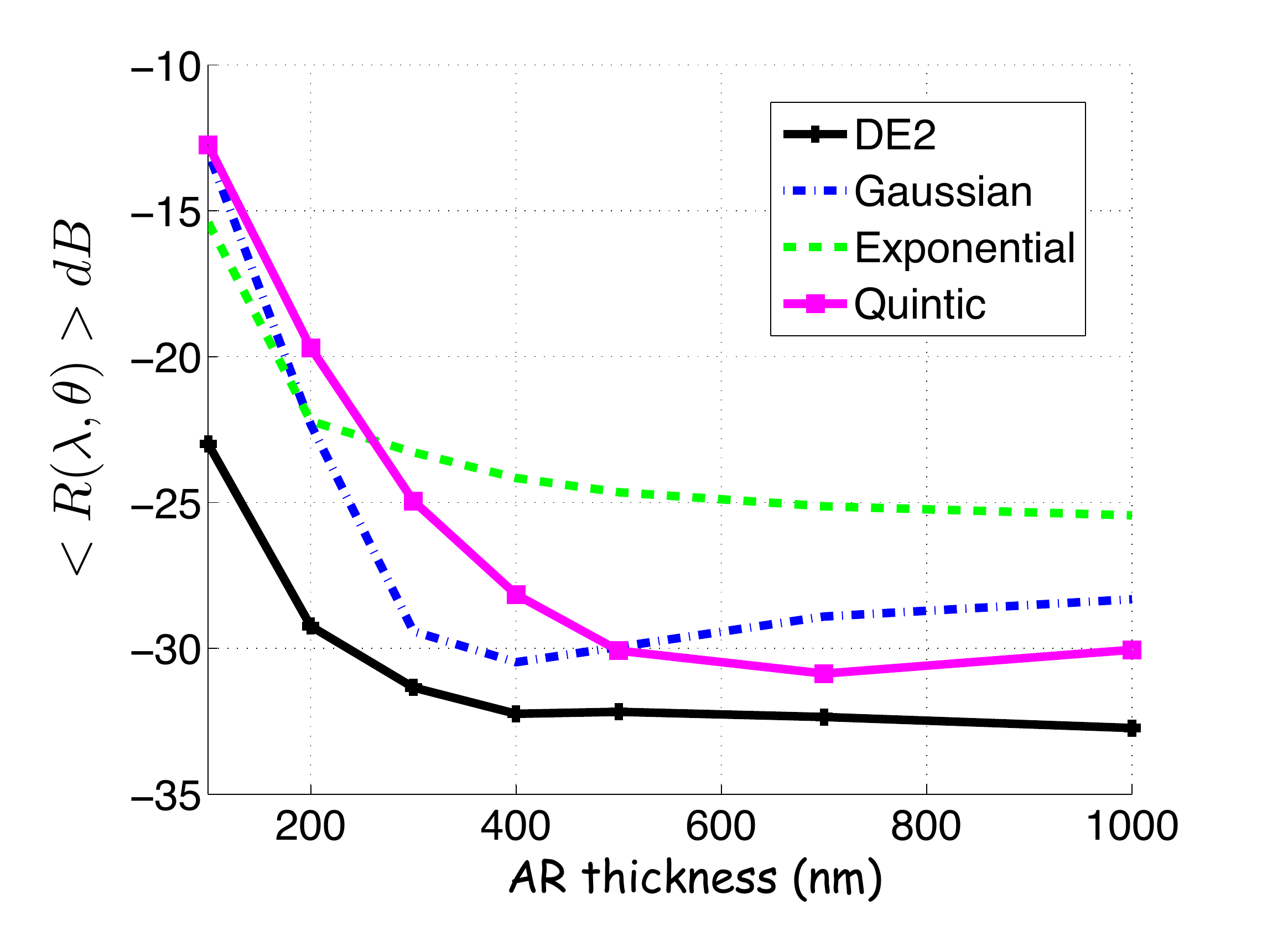}
\caption{Mean reflectivity plotted with a logarithm scale as a function of the AR coating thickness.} 
\end{figure}

To evaluate the gain in terms of the reduction of the reflectivity, we calculate the ratio of the average reflectivity for the Gaussian, exponential and quintic profiles to that of the optimized structures, Fig. 3. Very thin optimized structures with thicknesses of less than $300$ nm present a reflectivity at least twice lower than the conventional graded profiles. Figures 3b-c show the evolution of the RI profile for a thickness increasing from $100$ to $300$ nm for pixels of $20$nm. As explained above, thin structures of less than $300$ nm start with a first layer whose refractive index is higher than the minimal value of $1.2$. For a larger thickness the first hundred of nanometer is constituted of layers presenting the minimal refractive index $1.2$ just as for the graded profiles. For the last section of the structure, the refractive index increases toward the index of the silicon substrate but with some discontinuities. These brutal variations superimposed on a gradual pattern are clearly responsible for the enhancement of the efficiency of the optimal structures. This strategy explains that even thick AR coatings optimized by the DE2 algorithm overpass the Gaussian, Quintic or Exponential patterns. A thickness of $200 \, nm$ appears as a good compromise in terms of mean reflectivity ($3.4\%$) and reflectivity reduction (2.5 times lower than the Gaussian profile). This very thin optimized AR coating presents a mean reflectivity equivalent to that of a one micron thick gradual Gaussian or Quintic structure. The evolution of the reflectivity (averaged on both polarizations) for the $200 \, nm$ thick structure shown Fig. 3c is depicted on Fig. 4a. Its reflectivity averaged over the wavelength range $[400, 900] \, nm$ lowers to $0.15\%$ in normal incidence, reaches $0.6\%$ at $45^\circ$ and increases to $2.9\%$ at $60^\circ$. 

\begin{figure}[htbp]
\centering
\includegraphics[width=\linewidth]{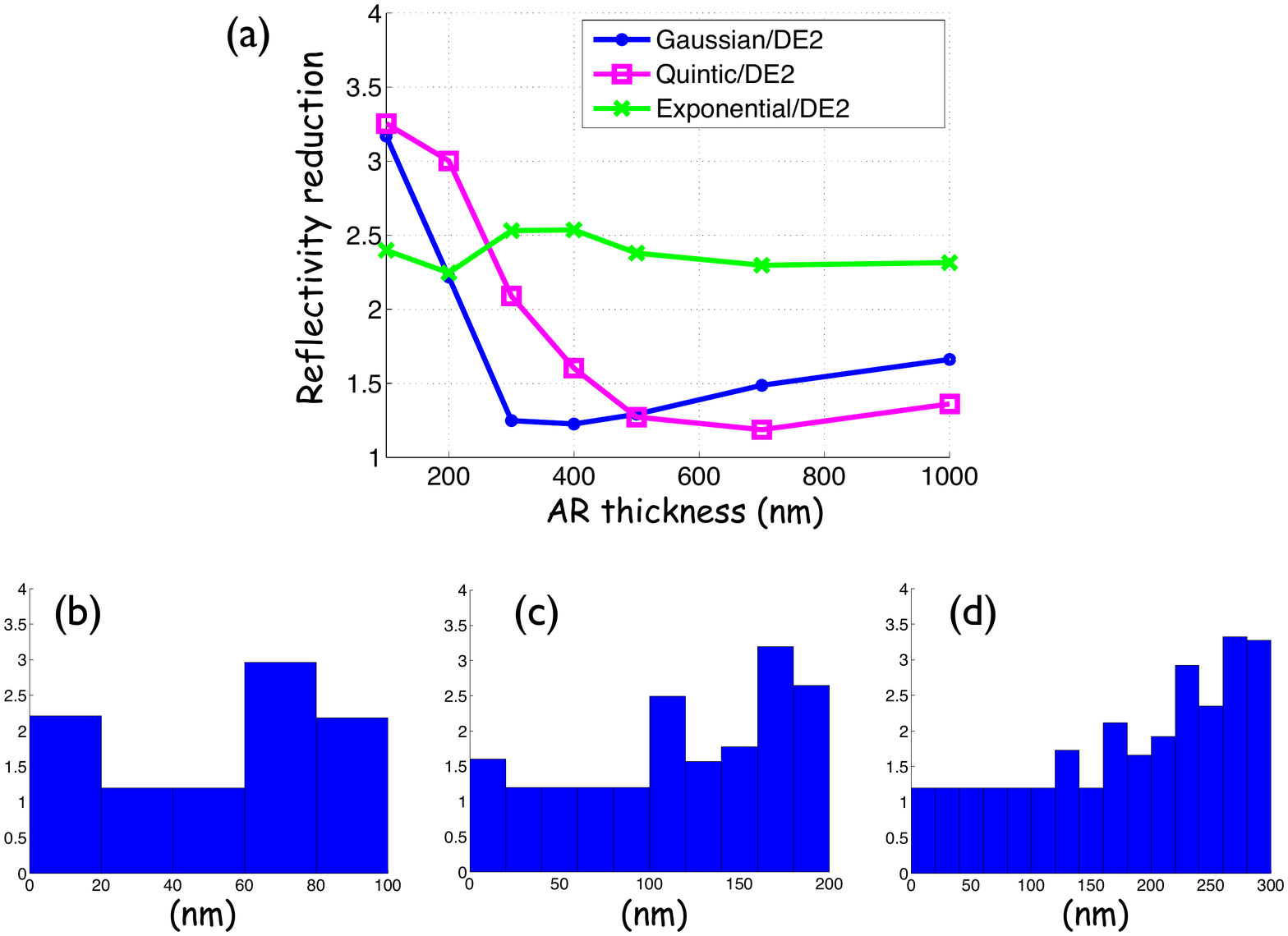}
\caption{ (a) Reflectivity reduction as a function of the thickness $d$. (b,c and d) RI profiles for the optimized AR coatings of respective thicknesses $100, 200, 300 \, (nm)$.} 
\label{fig3}
\end{figure}

The robustness of the $200 \, nm$ thick optimal structure against errors in the RI profile is evaluated  on  figure 4b. For that purpose, the mean reflectivity  $<R(\lambda, \theta>$ is averaged over $1000$ trials for which a random modulation of strength $\Delta n/n$ is applied to  the RI of each the 10 pixels. Although the mean reflectivity is degraded as the RI errors increase, it reaches only $4.1\%$ when each pixel is affected by a high RI variation of  $\Delta n/n=10\%$. These results prove that efficient and robust ultra thin AR coating can be realized with the current fabrication processes.  

To demonstrate how effective is our approach, we design a {\em realistic} $200$ nm thick AR coatings consisting of 4-layers on top of a a thick crystalline silicon wafer.  The DE2 algorithm is used to find the optimal AR coating for such an overall thickness. The refractive index and the thickness of each layer are the parameters that have to be optimized. They  were respectively restricted to the intervals  $[20,100]$ nm and  $[1.2, 3.5]$, by considering only lossless and non dispersive materials. The optical parameters of our optimal AR coating, shown figure 5a, are  $[24, 61, 72, 43]$ nm for the thicknesses  and $[1.6,1.2,2.2,3.4]$ for the refractive indices. 
The theoretical optical properties of this 4-layers structure evaluated with the optical absorption for the silicon wafer are very close to that obtained with the 10 pixels AR coating. The average reflectivity is $0.22\%$, $1.1\%$ and $4.1\%$ respectively for normal incidence, a $45^\circ$ and a $60^\circ$ incidence angle. The mean reflectivity $<R(\lambda, \theta)>$  slightly increases to $4.3 \%$ when the losses of the silicon are considered.

As illustrated by our approach, effective AR coatings require a fine control of each layer's refractive index and thickness. The stoechiometric materials available indices' list is finite and not precise enough to realize such simulated structures. Moreover the GLAncing Angle Deposition method (GLAD) is not adapted for structures that do not present a monotonic increase of the refractive index conversely to gradual AR coatings \cite{Schubert:2008uj,Kuo:2008p4393,Martin}. Our thin structures require indeed to deposit a high refractive index layer on top of a low refractive layer. We tackle this issue by the use of a reactive sputtering process. From an industrial point of view, sputtering is known to be easily scalable and the use of only one target reduces the fabrication cost of this system. The reactive sputtering process allowed us to realize the 4-layer AR coating and could also be used to realize more complex structures with a higher number of layers.

We deposit non-stoechiometric silicon oxynitride $Si_xO_yN_z$ layers thanks to reactive radiofrequency magnetron sputtering\cite{Farhaoui:2016hh}. With this technique, a pure silicon target (100 mm in diameter) is sputtered with a radiofrequency power at 250 W and 13.56 MHz in atmosphere of $Ar/O_2/N_2$. The substrate is placed on the rotating substrate holder at 9.5 cm in front of the target. The gas flows $F_{Ar}, F_{O2}$ and $F_{N2}$ are controlled by mass flowmeters. In this study, $F_{Ar}$ and $F_{N2}$ are fixed; whereas $O_2$ was periodically supplied using a rectangular pulsed flow rate from 0 during the off-time $T_{off}$, to a maximum flow rate $F_{max O2}$ during the on-time $T_{on}$. Depending on pulsed parameters ($T_{on}$, $T_{off}$ and $F_{max O2}$), time-averaged $O_2$ flow rate is tuned during target sputtering, and allows us to control precisely the composition of the deposited film, e.g. x, y and z in $Si_xO_yN_z$ material. Finally, we are able to obtain a large range of refractive indices from $1.45$ for silicon oxide to $3.74$ for silicon-rich films\cite{Farhaoui:2016hh}. 

\begin{figure}
\centering
\includegraphics[width=\linewidth]{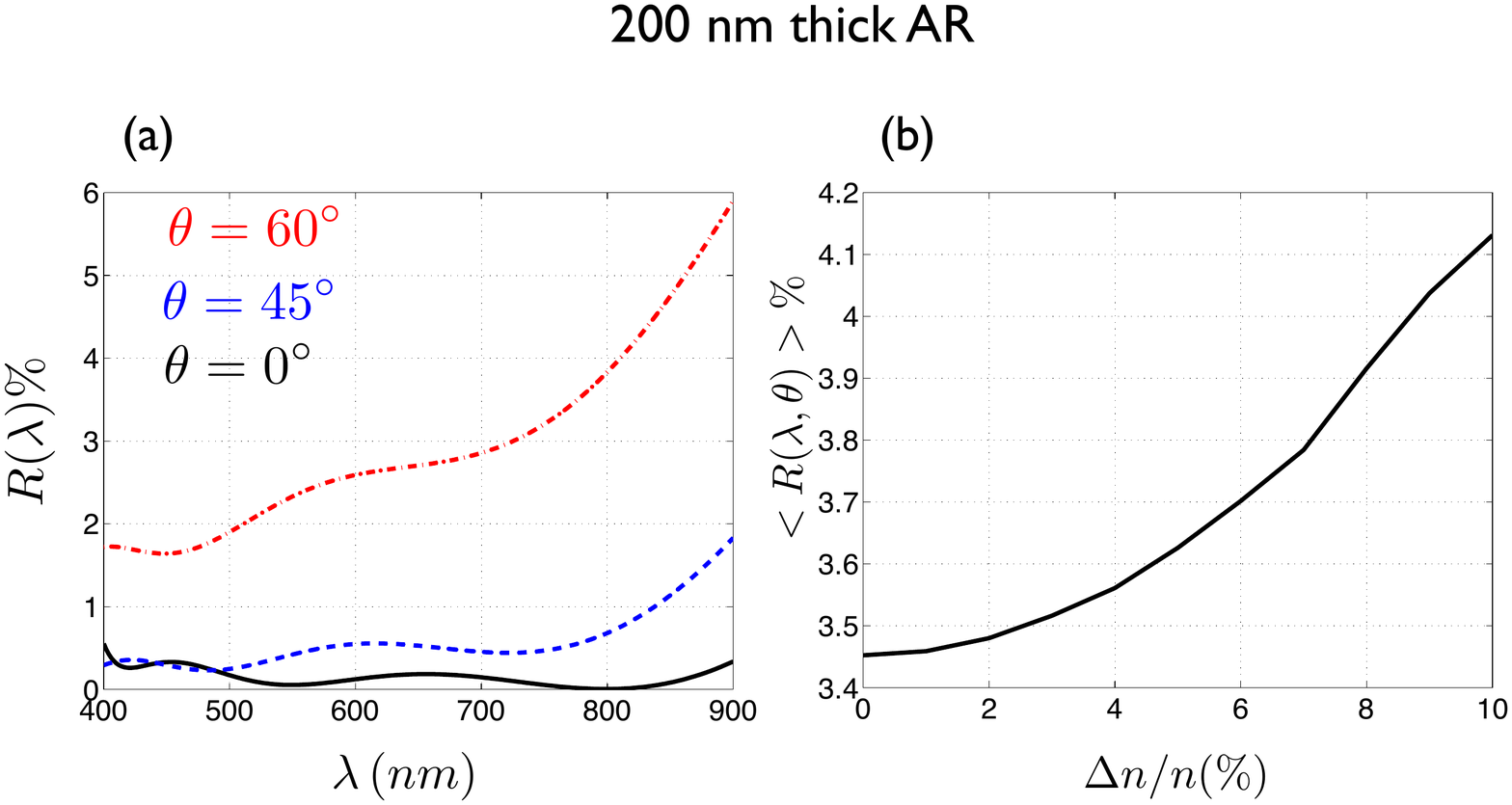}
\caption{(a) Reflectivity spectra calculated for three incident angles $0^\circ,45^\circ, 60^\circ$  for the optimized $200\, (nm)$ thick AR coating shown in fig. 3c. (b) Mean reflectivity as a function of the strength  $\Delta n/n$ of the RI disorder induced on each 10 pixels for a $200 \, nm$ AR coating.} 
\label{fig4}
\end{figure}

\begin{figure}[htbp]
\centering
\includegraphics[width=\linewidth]{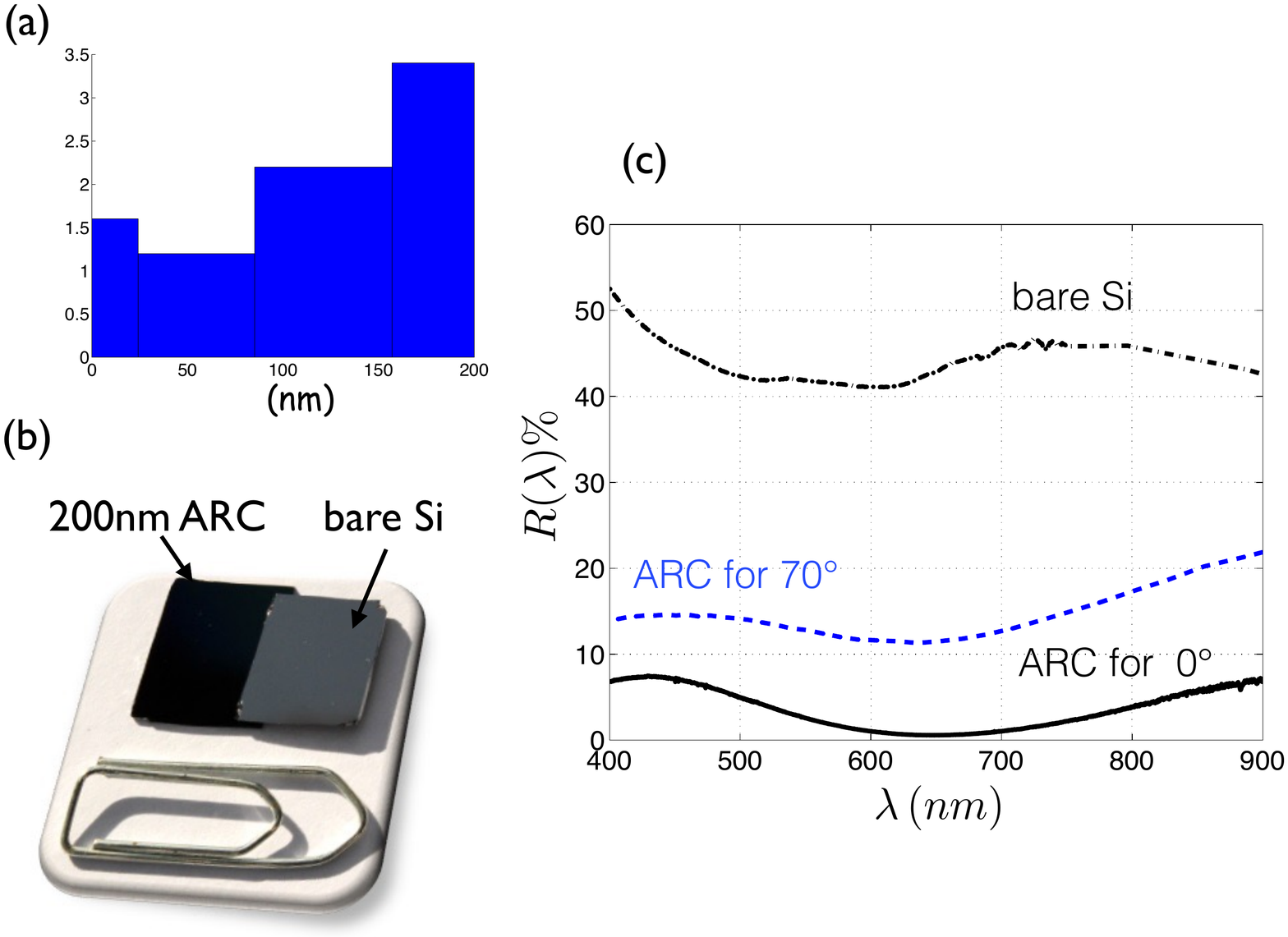}
\caption{ (a) RI profile for a 4 layers AR coating with an overall thickness of $200 \,nm$. (b) Photography of the silicon substrate with and without the 4 layer AR coating. (c) Experimental reflectivity as a function of the wavelength.} 
\label{fig5}
\end{figure}
The reflectivity of the structure shown in figure 5c is measured with a halogen lamp used as a visible light source. The reflected signal is imaged on the slits of a $32 \,cm$ focal length monochromator and then detected by a CCD camera. In normal incidence, the experimental spectrum exhibits a reflectivity between $0.56\%$ and $7.4\%$, Fig. 5c. In the $400-900 \,nm$ range, the experimental average reflectivity is reduced to $3.6\%$ with the AR coating instead of $44\%$ for the bare silicon wafer. For an oblique incident angle of $70^\circ$, the average reflectivity attains $14\%$. We underline that around $600 \,nm$, a wavelength for which the refractive index of the materials have been measured, the reflectivity is as low as could be expected. To reach even better performances for the realized device, the simulations should take into account the material dispersion and optical losses. This will be left for future works. However, our structure has a significant breakthrough in comparison with a simple quarter wavelength AR coating of $86 \,nm$ thick operating at $650 \,nm$ whose average reflectivity attains $14\%$ in normal incidence.
 
In conclusion, we have demonstrated that optimal thin AR coatings present a RI profile that combines a gradual envelop with an interferential pattern. The sub-wavelength patterning, presenting a clear interferential role, is shown to drastically improve the optical efficiency of AR coatings in terms of averaged reflectivity and angular acceptance. These non trivial RI discontinuous profiles designed by a evolutionary algorithm outperform the continuous Gaussian profile. These original structures have the advantage of being robust against fabrication errors. Finally, a 4 layers AR coating of  $200 \,nm$ thick has been realized using a vapor deposition technique. The experimental measurements  show a strong reduction of the reflectivity in normal incidence, thus demonstrating the feasibility of these nanostructured AR coatings with processes that can be easily deployed.

\section*{Funding Information}
The authors acknowledge the Labex IMOBS$^3$ for the financial support. 


\bibliography{biblio}

\end{document}